\documentclass[12pt]{iopart}
\usepackage{amssymb}
\expandafter\let\csname equation*\endcsname\relax
\expandafter\let\csname endequation*\endcsname\relax
\usepackage{amsmath}
\usepackage[english]{babel}
\usepackage{graphicx}
\usepackage{textcomp}
\usepackage{bibentry}

\newtheorem{remark}{Remark}%[section]
\newtheorem{proposition}{Proposition}%[section]
\begin{document}

\title[Stability of axisymmetric vortices in
compressible medium]{Nonlinear stability of two-dimensional
axisymmetric vortices in compressible inviscid  medium in
%non-inertial
a rotating reference frame}

\author{Olga S. Rozanova$^1$, Jui-Ling Yu$^{2,3}$,
Marko K. Turzynsky $^1$ and Chin-Kun Hu$^{4,5}$}

\address{$^1$Department of Mechanics and Mathematics, Moscow
State University, Moscow 119992 Russia}
\address{$^2$ Department of Financial and Computational Mathematics,
Providence University, Taichung, 43301, Taiwan}
\address{$^3$
 National Center
of Theoretical Sciences at Taipei, Physics Division, National Taiwan
University, Taipei 10617, Taiwan}
\address{$^4$Institute of Physics, Academia Sinica, Nankang, Taipei
11529, Taiwan}
\address{$^5$National Center  for   Theoretical   Sciences,
National Tsing   Hua   University,  Hsinchu  30013,   Taiwan}

\ead{rozanova@mech.math.msu.su}

\begin{abstract}
We study the stability of the vortex in a 2D model of continuous
compressible media in a uniformly rotating reference frame.  As it
is known, the axisymmetric
 vortex in a fixed reference frame  is stable with
respect to asymmetric perturbations for the solution of the 2D
incompressible Euler equations and basically instable for
compressible Euler equations. We show that the situation is quite
different for a compressible axisymmetric
 vortex in a rotating reference
frame. First, we consider special solutions with linear profile of
velocity (or with spatially-uniform velocity gradients), which are
important because many real vortices have similar structure near
their centers. We analyze both cyclonic and anticyclonic cases and
show that the stability of the solution depends only on the ratio of
the vorticity to the Coriolis parameter. Using a very delicate
analysis along with computer aided proof, we show that the stability
of solutions can  take place only for a narrow range of this ratio.
 Our results imply that the rotation of the coordinate frame can
stabilize the compressible vortex. Further, we perform both
analytical and numerical analysis of stability for real-shaped
vortices.
\end{abstract}


\vspace{2pc}


% \pacs{92.60Aa,47.10ad, 92.60Pw, 86A10}

\submitto{\NL}


\maketitle
\nocite{*}
% For two-column output uncomment the next line and choose [10pt] rather than [12pt] in the\documentclass declaration

%\end{document}

Many flows in oceans and atmospheres are approximately
two-dimensional. The vortex structures are their characteristic
feature. Two-dimensional vortices also play an important role in
tokamak-confined plasmas \cite{plasma} as well as in astrophysical
situations such as accretion discs of neutron stars \cite{stars}.
Although the vortex dynamics can be complicated, it is natural to
begin with the study of certain elementary processes. One example is
the stability of  isolated circular free vortices in rotating
fluids. Their  stability/instability properties are of fundamental
interest for refined models of general atmospheric circulation due
to the presence of strain and shear in the ambient flow.

%Concerning the atmospherical model we can add that

%with respect to asymmetric perturbations.

%In this paper we consider vortex motion in inviscid compressible
%media.
The vortex motion is traditionally considered in incompressible
media. The theory of such motion, going back to the classical works
of Helmholtz, Gr\"obli, Kirchhoff, Rankin, Greenhill, Taylor,
Poincar\'e, has a huge literature (see, e.g. \cite{VortexBibl},  for
a modern state-of-art \cite{Flor1}, \cite{Flor2}, \cite{Dolzhansky}
). Stability of the vortex is the most classical issue. It can be
understood in many different ways \cite{Leblanc}, \cite{Flor2},
\cite{Sipp}, \cite{Flierl}, \cite{Kloosterziel}, \cite{Hopfinger}.
 Numerical
models demonstrate that there is a broad class of geophysical
vortices freely evolve toward axisymmetric states. This intrinsic
drive toward symmetry opposes the destructive shearing from the
environmental flow \cite{Schecter1}. Moreover, as it is shown in
\cite{Inviscid_damping}, the axisymmetric form of the vortex is
stable with respect to asymmetric perturbations for the
two-dimensional inviscid flow.

Nevertheless, the vortex motion in compressible media is also very
important both theoretically and practically. Let us show this by
example. It is well known that the processes of convergence and
divergence of the air flows play an important role in the atmosphere
dynamics. Therefore, if we average over the height the original
tree-dimensional system of equations for dynamics of the air (using
the fact that the horizontal scale is much larger than the vertical
one), the resulting two-dimensional model should be compressible in
order to reflect the processes.

The earliest results concerning vortices in the compressible fluid
were obtained by Chree \cite{Chree} in the meteorological context. A
good review of further works on 2D compressible vortices emphasizing
experimental aspects can be found in \cite{Bershader}. In
\cite{Colonius}, authors present analytical solutions for the
reduced system of equations corresponding to free compressible
viscous vortices and compare their results with numerical solutions
of the full system of equations as well as experiments. In
\cite{Lu}, \cite{Caillo}, authors discuss different situation of
instability in an inviscid compressible media. There exist works,
where a linear analysis of stability of compressible vortex was
performed in various setting, see \cite{Fung}, \cite{Khorrami},
\cite{Rusak},\cite{Chan}, \cite{Men'shov}, \cite{Hiejima} and
references therein.
 In \cite{Ardalan},
\cite{Chiocchia}, \cite{Aboelkassem}  steady compressible vortices
were constructed using hodograph plane transformations. Mathematical
analysis of self-similar isentropic non-rotational 2D Euler system
including shock wave structure can be found in \cite{Zheng_chapter},
see also references therein. In particular, there a non-steady
solution corresponding to axisymmetric vortex was obtained.

In the present paper we study the effects of compressibility for a
special class of solution of the Euler equations given on a plane
rotating with a constant angular velocity. This is a class of
solution with spatially-uniform velocity gradients known from
Kirchhoff. This class of solution has many outstanding applications,
for example, the flow of an ideal homogeneous incompressible fluid
inside a triaxial ellipsoid \cite{Lamb}, the theory of the motions
of solids with cavities filled with a fluid, and the theory of fluid
gyroscopes in the field of Coriolis forces \cite{Dolzhansky}.

In the incompressible case the Kirchhoff vortex is known to be an
analytical solution of the two-dimensional Euler equations (see
\cite{Lamb}). This vortex shows a steady rotation without changing
its shape. This solution  provides a possibility for constructing
the vortex patch\footnote{ The vortex patches exist in the 2D
compressible media, too \cite{secchi}}.

We consider the solution with spatially-uniform velocity gradients
for isentropic compressible Euler equations in a rotating coordinate
frame. In this case, the only possible "Kirchhoff vortex" with
constant vorticity $\omega$ is axisymmetric vortex.  Its nonlinear
stability within the class of asymmetric motion having
spatially-uniform velocity gradients is under consideration. The
elliptical patch does not demonstrate solid body rotation anymore:
behavior of the trajectories  becomes very complicated.

 As in the theory of fluid gyroscopes, the problem is reduced to a study
of stability of equilibrium for a quadratically nonlinear system of
ordinary differential equations in a phase space of large
dimensions. In particular, a linear analysis shows that on some
range of parameter $\omega/l$, where $l$ is the Coriolis parameter,
even small break of axial symmetry in the initial data leads to
break of stability. The linear analysis does not give an answer for
the rest of parameters.  The  analysis of nonlinear stability  is
much more complicated. In fact, we prove that in general, (except of
a discrete number of resonant parameters) the motion is
quasi-periodic and therefore stable in the Lyapunov sense up to the
the third order terms in the respective normal form. In other words,
we prove the necessary stability condition for the "stable" region.
This "stable" region includes  both cases of cyclonic (low pressure
in the center of vortex) and anticyclonic (high pressure in the
center of vortex) motions. Moreover, the cyclonic motion can be
anticlockwise (normal low) or clockwise (anomalous low). If the
parameter $\omega/l$ approaches to the very boundaries of linear
unstable cyclonic region, the necessary condition for stability
fails, and the resonant values appear in a large amount in the
domain of anticlockwise motion. Also, the necessary condition for
stability fails when the cyclonic values of parameters switch to the
cyclonic values and vice versa. The anticyclonic domain contains
many resonant values of the parameter, nevertheless, at other points
the necessary condition for stability holds.

The problem considered seems artificial since the structure of real
vortices in fluid is much more complicated  than we discussed here.
Indeed, as it is known from laboratory experiments and from
observation of large atmospheric vortices like tropical cyclone
\cite{IAV},  the velocity has linear profile near the center of
vortex. Moreover, as follows from experiments with 2D transverse
compressible vortices, the swirl profile defines a vortex  core that
rotates  almost as  a rigid body and decays far from the center.
This finding is generally consistent with profiles of incompressible
vortices. However, the stability of vortex depends on the spatial
decay rate
 for circular vortex in incompressible
fluids (see modified Rayleigh criterion \cite{Flor1}).

We performed a series of numerical experiments with initial data
corresponding to elliptical deformation of exact axially symmetric
stationary solution to 2D inviscid compressible Euler equations
constructed in our previous work \cite{RYH2012}. This solution has
exponential spatial decay at infinity and was used for modeling a
moving tropical cyclone in averaged 2D atmosphere.

In these numerical experiments we  investigate the stability of
cyclonic and anticyclonic vortices with spatially decaying "tails"
and the relations with  theoretical results on stability of the
vortices having linear velocity profile. We show that for
sufficiently small vorticity, the level lines of the pressure near
the center of vortex are approximately elliptic. If the vorticity
increases, the picture of level lines is similar to the azimuthal
Kelvin--Helmholtz instability as observed by Chomaz et al., see
 \cite{Flor2}. If the vorticity increases further, the numerical solution demonstrates
 the formation of shock waves.

%The nonlinear stability of a steady basic flow for inviscid
%barotropic gases with respect to the class of smooth perturbations
%with suitable boundary condition can be proved by the
%Lagrange-Dirichlet method \cite{Padula}.

The paper is organized as follows. In Sec.\ref{Sec2}, we introduce a
two-dimensional system of PDEs, which follows from  compressible
isentropic inviscid Euler equations after doing appropriate change
of variables. In Sec.\ref{Exact_sol}, we introduce a special class
of solutions for this system and obtain the system of ODEs governing
its behavior. In Sec.\ref{finite_mass}, we consider another class of
solutions characterized by conservation of mass and energy and show
that these solutions are governed by the same system of ODEs. Then
we study the stability of the unique equilibrium of this system that
refers to a circular steady vortex. In Sec.\ref{s1}, we analyze some
simplification of the full system, where the system can be exactly
integrated.  In Sec.\ref{general_case}, we consider the full system.
The linearization at equilibrium gives us the instability domain,
however, for the rest of parameters the matrix of linearization has
three couples of pure imaginary complex conjugate eigenvalues.
Therefore, the linear theory does not answer whether the equilibrium
is stable or not. To study nonlinear stability, we use the theory of
normal forms and apply the theorem about existence of almost
periodic trajectories. Due to the difficulty of algebraic
computations, we perform them by the computer algebra system MAPLE.
The normal forms are analyzed up to the third order. This gives us a
possibility to conclude that the system is unstable both near the
boundaries of instability domain and near the points where the
cyclonic motion changes to anticyclonic one. For the rest of points,
may be except for some discrete values, the necessary condition for
the stability holds. Moreover, up to the third order term in the
normal form expansion, the sufficient condition for stability holds,
too. In Sec.\ref{Sec3}, we present the results of numeric
experiments for localized vortices having  linear velocity near
their center and spatially decaying at infinity. These experiments
show that for
 sufficiently small vorticity the the results about stability for the special class of solutions with
linear profile of velocity can catch the main stability features of
real-shaped localized vortex. Moreover, we show analytically the
reason why the domains of stability for the real-shaped localized
vortex basically cannot coincide with the domain of stability for
the vortex with linear profile of velocity for large values of
vorticity. In Sec.\ref{Sec4}, we summarize our results.
% and outline prospects
%for further researches.

\section{Bidimensional models of rotating compressible medium}\label{Sec2}

The two-dimensional system of motion of inviscous compressible
medium on a rotating plane consists of three equations for density
$\varrho(t,x),$ velocity ${\bf U}(t,x)$ and pressure $p(t, x)$:
\begin{equation}\label{2d_U}
%\begin{split}
 \varrho(\partial_t {\bf U} +  ({\bf U}\cdot \nabla ){\bf U} +
 {\mathcal L}{\bf U}) + \nabla p  = 0,
 \end{equation}
 %\end{split}
%\end{align}
%\end{equation}
%\begin{equation}
%\begin{align}
%\begin{split}
\begin{equation}\label{2d_rho}
\partial_t \varrho +  {\rm div} ( \varrho {\bf U}) =0,
\end{equation}
% \end{split}
%\end{align}
%\end{equation}
%\begin{equation}
%\begin{align}
%\begin{split}
\begin{equation}
\partial_t p + ({\bf U}\cdot\nabla p)+\gamma p\,{\rm div}{\bf
U}=0.\label{2d_p}
% \end{split}
%\end{align}
\end{equation}
Here $\mathcal L = l L $,
$\quad L = \left(\begin{array}{cr} 0 & -1 \\
1 & 0
\end{array}\right)$,
 $\gamma\in (1,2)$ is the heat ratio,
$l>0$ is the Coriolis parameter. $\nabla$ and $\bf\rm  div$ denote
the gradient and divergence with respect to the space variables.
$(.\,\cdot \, .)$ is the inner product. Under suitable boundary
conditions, the system \eqref{2d_U} -- \eqref{2d_p} implies
conservation  of mass, momentum and total energy.

%\end{document}

If we restrict ourselves to the barotropic case, where $p=C
\rho^\gamma,\,  C={\rm const}$, then the system under consideration
can be reduced to two equations \eqref{2d_U}, \eqref{2d_rho}.

In our previous papers \cite{RYH2010}, \cite{RYH2012}, we studied
this model in the context of atmosphere dynamics.  We started from
the (primitive) three-dimensional system of equations for
compressible rotating Newtonian polytropic gas \cite{Landau},
\cite{Pedloski} and apply the  procedure of averaging over the
height to obtain a two-dimensional system of equations (see
\cite{Obukhov}). The vortices in this model were associated with
typhoons. In this paper, we do not specify the two-dimensional
compressible medium. The results can be applied to other model,
nevertheless, our main interest is a possibility of existence of
long living atmospherical vortices, therefore in numerical
experiments we use the parameters related to the atmosphere.

It will be convenient for us to introduce a new variable
$\Pi=p^{\frac{\gamma-1}{\gamma}}$ and reduce \eqref{2d_U},
\eqref{2d_rho} to
\begin{equation} \label{main}
\partial_t {\bf U} + ({\bf U} \cdot \nabla ){\bf U} +  {\mathcal L}
\,{\bf U} + c_0\, \nabla \Pi  =0,
\end{equation}
\begin{equation}\label{pi}
\partial_t \Pi +  (\nabla \Pi \cdot {\bf U}) +
(\gamma-1)\,\Pi\,{\rm div}\,{\bf U}=\,0, \end{equation} $c_0=
\frac{\gamma}{\gamma-1} C^{\frac{1}{\gamma}}.$
%In what follows we
%deal with system \eqref{main}, \eqref{pi}.

\section{A class of exact solutions}\label{Exact_sol}

Let us look for the solution of \eqref{main}, \eqref{pi} in the form
\begin{equation}\label{u-form}{\bf U}(t,{\bf x})=Q {\bf x},\qquad Q= \left(\begin{array}{cr} a(t) & b(t) \\
c(t)& d(t)\end{array}\right),\end{equation}
\begin{equation}\label{pi-form}
\Pi(t,{\bf
x})=A(t)x_1^2+B(t)x_1x_2+C(t)x_2^2 +K(t).% \end{aligned}
\end{equation}
In fact, the solution is the first term of the Taylor series
expansion at the point of minimum or maximum of pressure: the first
and second term represent velocity and pressure respectively. In
this way, we can keep a maximum possible members in this expansion
to obtain an exact solution of \eqref{main} and \eqref{pi}.

Thus, we get a closed ODE system for the components of the matrices
$Q$
and $R=\left(\begin{array}{cr} A(t) & \frac12 B(t) \\
\frac12 B(t)&C(t),\end{array}\right) $:
\begin{equation}
%\begin{aligned}
\label{m1}
\dot R+ RQ + Q^T R+(\gamma-1){\bf tr}Q R =0,\end{equation}
\begin{equation}
\label{m2} \dot Q+Q^2+lLQ+2c_0 R=0,
\end{equation}
\begin{equation*}
\dot K+2(\gamma-1){\bf tr} Q K=0.\label{K}
% \end{aligned}
\end{equation*}
%where
%$ T= \left(\begin{array}{cr} 0 & -1 \\
%1 & 0\end{array}\right).$
The last equation is linear with respect to $K$, whereas
 the system
of matrix equations \eqref{m1},  \eqref{m2} consists of 7 nonlinear
ODEs and has a very complicated behavior.

In fact, an analogous class of solutions was considered in
\cite{Ball} in another context (see also \cite{Thacker}).

\section{Solutions with a finite mass}\label{finite_mass}

System \eqref{m1},  \eqref{m2} can also be applied to describe a
behavior of a finite mass of compressible media in non-polytropic
case (see \cite{Roz_nova} and references therein).

Let us recall briefly the derivation. For smooth solutions to
\eqref{2d_U}-\eqref{2d_p}, we consider the integrals of total mass
${\cal M}=\int\limits_{{\mathbb  R}^2} \rho d{x}$ and total energy
${{\cal E}}(t)=\int\limits_{{\mathbb
 R}^2} \left( \frac{\rho |{\bf
U}|^2)}{2}+\frac{p}{\gamma-1} \right) \,d{x}={\cal E}_k(t)+{\cal
E}_p(t)$.   To guarantee the convergence of the integrals, we can
assume that the density and pressure vanish as $|{\bf x}|\to\infty$
rather quickly, whereas the velocity components may even grow. For
these solutions, the total mass and total energy are conserved.
Further, we introduce the following
 functionals: $$G(t)=\frac{1}{2}\int \limits_{{\mathbb R}^2}\rho|{\bf
x}|^2\,d{ x},\quad  F_i(t)=\int\limits_{{\mathbb
 R}^2}({\bf U, X}_i)\rho\,d{ x}$$ and
$$
G_{x_1}(t)=\frac{1}{2}\int\limits_{{\mathbb R}^2} \rho x_1^2 d{
x},\quad G_{x_2}(t)=\frac{1}{2}\int\limits_{{\mathbb R}^2} \rho
x_2^2 d{ x}, \quad G_{x_1 x_2}(t)=\frac{1}{2}\int\limits_{{\mathbb
R}^2} \rho x_1 x_2 \,d{ x},
$$
  where ${\bf X}_1=(x_1,x_2),\,{\bf X}_2={\bf
r}_\bot=(x_2,-x_1),$ $\, i=1,2$. We note that for nontrivial
solutions $G(t)>0$ and $\Delta(t)=G_{x_1}(t)G_{x_2}(t)-G_{x_1
x_2}^2(t)>0$.
 It
can be readily checked by means of the Green's formula that for the
classical solutions to \eqref{2d_U}-\eqref{2d_p} the following
relations hold:
\begin{equation*}%\label(2.1.1)
G'(t)=F_1(t),\quad \quad F_2'(t)=l F_1(t)
\end{equation*}
\begin{equation*}
F_1'(t)=2(\gamma-1){\mathcal E}_p(t)+2 {\mathcal E}_k(t) - l
F_2(t),\end{equation*}
%\begin{equation}\label(2.1.7){\cal
%E}'(t)= 0.
%\end{equation}

Moreover, for the velocity \eqref{u-form} we have
$$G_{x_1}'=2a G_{x_1}+2bG_{x_1 x_2},\quad G_{x_2}'=2d G_{x_2}+2cG_{x_1 x_2},
\quad G_{x_1 x_2}'=(a+d) G_{x_1 x_2}+ b G_{x_2}+ c G_{x_1},
$$
$$
E_p'(t)=-(\gamma-1)(a(t)+d(t))E_p(t), \quad
\Delta'(t)=2(a(t)+d(t))\Delta(t).
$$
the potential energy $E_p(t)$ is connected with $\Delta(t)$ as
follows:
$$E_p(t)=E_p(0)\Delta^{(\gamma-1)/2}(0)\Delta^{(-\gamma+1)/2}(t).$$
Let us introduce new functions
$$G_1(t)={G_{x_1}(t)}{\Delta^{-(\gamma+1)/2}(t)},\quad
G_2(t)={G_{x_2}(t)}{\Delta^{-(\gamma+1)/2}(t)},\quad G_3(t)={G_{x_1
x_2}}{\Delta^{-(\gamma+1)/2}(t)}.$$ For the elements of the matrix
$Q$ and $G_1,\,G_2,\,G_3$ we get the closed system of equations
$$
a'=-a^2-bc+lc-\mathcal K G_2,
$$
$$
b'=-b(a+d)+ld+\mathcal K G_3, $$
$$
c'=-c(a+d)-la+\mathcal KG_3, $$
$$
d'=-d^2-bc-lb-\mathcal K G_1, $$
$$
G_1'=((1-\gamma)a-(1+\gamma)d)G_1+2bG_3, $$
$$
G_2'=((1-\gamma)d-(1+\gamma)a)G_2+2cG_3,
$$
$$
G_3'=cG_1+bG_2-\gamma(a+d)G_3,
$$
 with
$\mathcal K=\frac{\gamma-1}{2} E_p(0)\Delta^{(\gamma-1)/2}(0).$ The
system coincides with \eqref{m1}, \eqref{m2}, where $2 c_0=\mathcal
K$, $A=G_2$, $B=-2G_3$, $C=G_1$. Nevertheless, in this case the
conservation of the total energy $\mathcal E$ implies a
supplementary first integral.  Moreover, for $\gamma=2$, the
explicit form of $F_1$ and $F_2$ allows to obtain non-autonomous
first integrals.

\subsection{The case without non inertial forces}

In the case $l=0$, the problem about expanding a finite mass of gas
in vacuum was studied extendedly in Lagrangian coordinates
$\alpha=(\alpha_1,\alpha_2)$, where $x(\alpha,t) = {\mathcal
F}(t)\alpha$, ${\mathcal F}(t)$ is a square matrix. Thus, $U(x,t) =
\dot x = \dot {\mathcal F}(t){\mathcal F}^{-1}(t)x$ and $Q(t)=\dot
{\mathcal F}(t){\mathcal F}^{-1}(t)$ following from \eqref{u-form}.
In \cite{Ovsyannikov}, it was shown that the system of gas dynamics
with linear profile  of velocity for a polytropic gas in the $n$-
dimensional space can be reduced to a system of $n^2$ ODE of the
second order
\begin{equation}\label{mathcalF}
{\mathcal F}^T \ddot{\mathcal F} + ({\rm det} {\mathcal F})^{
1-\gamma} {\mathcal V} = 0,
\end{equation}
where  ${\mathcal V}$ is a constant matrix. The theory of these
equations was developed in \cite{Dyson}, where first integrals
correspond to the conservation of angular momentum, and the
vorticity were found. After that, in \cite{Anisimov_Lysikov}, the
authors found a supplementary non-autonomous first integral that
allow to prove the integrability of system \eqref{mathcalF} for
$n=2$ and $\gamma=2$ for the shallow water case. Bogoyavlensky in
\cite{Bogoyavlensky} found several general properties of dynamics of
ellipsoid with uniform deformation for $n=3$. For example, he
estimated the growth of the sum of squares of the main axis of the
ellipsoid. In \cite{Anisimov_Inogamov}, it was proved that the 3D
gas ellipsoid is unstable and tends to shrink into a plane ellipse.
This result was known before from numerical computations. The cases
of integrability in three dimensional case were found by Gaffet
\cite{Gaffet}.

There exists a huge literature concerning the analogs of the problem
for the incompressible liquid, the self-gravitating gas including
the stability issue, and the Hamiltonian formulation of the problem
for both incompressible and compressible cases. We refer to the book
\cite{Bogoyavlensky} and the recent review papers
\cite{Borisov_Mamaev_Kilin}, see also the references therein.

%\subsection{A friction-free vortex ($\mu=0$), stability issue}

\section{Axisymmetric case \cite{RYH2010}}\label{s1}

 It is easy to
see that system \eqref{m1},  \eqref{m2} has a closed submanifold of
solutions with additional properties $a=d$, $c=-b$, $A=C$, $B=0$.
These solutions corresponds to the axisymmetric motion. Note that it
is the most interesting case related to the vortex in atmosphere.
Here we get a system of 3 ODEs:
\begin{equation}\begin{aligned}\label{As0}\dot A+2\gamma
aA=0,\\ \dot a+a^2-b^2+lb+2c_0 A=0, \\
%\label{as0}
\dot b+2ab-la=0.%\label{bs0}
\end{aligned}
\end{equation}

The functions $a,\,b,\,A$ correspond to one half of divergence, one
half of vorticity and the fall or rise of pressure in the center of
vortex respectively. The only nontrivial equilibrium point that
relates to a vortex motion is
\begin{equation}\label{equililria}
a=0,\, b=-c=b^*,\, A=A^*=\frac{b^*(b^*-l)}{2c_0}.\end{equation}
 If
$A^*>0$, then the center of vortex corresponds to a domain of low
pressure  (the motion is cyclonic). This implies $b^*<0$ or $b^*>l$.
If $A^*<0$, that is $0<b^*<l$, the motion is anticyclonic. The
degenerate case $A^*=0$ of constant pressure corresponds to the case
$b^*=0$ or $b^*=l$.

 Further, if $A\ne 0$, then there exists one
first integral
\begin{equation*}\label{mu0_integral}
b(t)=\frac {l}{2}+ \mathcal
C|A(t)|^{\frac{1}{\gamma}},\end{equation*} where $C$ is a constant
\cite{RYH2010}. Thus, \eqref{As0} can be reduced to the following
system:
\begin{equation*}%\begin{aligned}
\label{A2s} \dot A(t)=-2\gamma a A, \quad \dot
a(t)=-a^{2}-\frac{l^{2}}{4}+\mathcal C^{2} A^{\frac{2}{\gamma}}-2
c_{0} A.
%\label{a2s}
%\end{aligned}
\end{equation*}
On the phase plane $(A,a)$, there always exists a unique
 equilibrium  $(A^{*},a^{*})=(A_{0},0)$, stable in the Lyapunov sense
(a center), where $A_{0}$ is a  root of equation
$$\frac{l^{2}}{4}+2c_{0}A=\mathcal C^{2}A^{\frac{2}{\gamma}}.$$

Let us notice that if $\mathcal C=0$, we have a particular case of
anticyclonic motion with a constant vorticity $2b=l$.

 If $A=0$, then \eqref{As0} can be reduced to one
Riccati equation \begin{equation*}%\begin{aligned}
\label{z}\dot z=-z^{2}+il z\end{equation*} for the function
$z(t)=a(t)+i b(t): {\mathbb C}\mapsto {\mathbb C}$. It has an
explicit solution $z=\frac{lz(0)}{(l+iz(0))e^{-il t}-Iz(0)}$, $a=\Re
z$, $b=\Im z$.

\begin{remark}
In \cite{Zheng97}, \cite{Zheng98} (see also \cite{Zheng_chapter})
for the case of absence of a non-inertial force  the authors
construct construct a two-parameter family of self-similar
non-steady solutions to the compressible two-dimensional Euler
equations with axial symmetry. The equations can be reduced to two
systems of ordinary differential equations. In the  polytropic case,
the system in autonomous form consists of four ordinary differential
equations with a two-dimensional set of stationary points, one of
which is degenerate up to order four. Through asymptotic analysis
and computations of numerical solutions, the authors recognize a
one-parameter family of exact solutions in explicit form
corresponding to a vortex. All the solutions (exact or numerical)
are globally bounded and continuous, have finite local energy and
vorticity, and have well-defined initial and boundary values at time
zero and spatial infinity respectively. Particle trajectories of
some of these solutions are spiral-like. Near the center of vortex
the velocity has a linear profile.
\end{remark}

\section{General case}\label{general_case}

As follows from \cite{Inviscid_damping}, the axisymmetric form of 2D
vortex is exponentially stable with respect to asymmetric
perturbations for the solution to the incompressible Euler
equations. Indeed, the incompressibility condition implies
$a(t)+d(t)=0$ and this reduces the full system \eqref{m1},
\eqref{m2} to \eqref{As0}. As we have shown in Sec.\ref{s1}, the
equilibrium in this case  is stable for any $b^*$ and $l$.

Nevertheless, in the compressible case  this property does not hold
for arbitrary values of parameters.

As one can check, the point
\begin{equation}\label{equililriaf}
a=d=0,\, b=-c=b^*,\, A=C=A*=\frac{b^*(b^*-l)}{2c_0},\, B=0
\end{equation}
 is the
only equilibrium of the full system \eqref{m1}. It is the same point
of equilibrium \eqref{equililria} as in the axisymmetric case
\eqref{As0}. Nevertheless, in the symmetric case this equilibrium is
always stable in the Lyapunov sense, whereas in the general case the
situation is different.

The direct computation shows that the following property holds.
\begin{proposition}\label{prop_first_int}
 The  system \eqref{m1} has the first integral
\begin{equation}\label{first_int}
b-c-l= {\rm const}\,{\mathcal D}^{\frac{1}{2\gamma}},\quad {\mathcal
D}=4AC-B^2.
\end{equation}
\end{proposition}
This reduces \eqref{m1}, \eqref{m2} to the system of 6 equations,
equation for $b(t)$ will be excluded.
%To our knowledge, any further
%reduction of dimension is not possible in the general case.

Let us notice that the case $c=b-l$ is exceptional, since the
constant in \eqref{first_int} is equal to zero and the relation
between $c$ and $b$ is very simple. This case can be called the case
of zero potential vorticity (e.g. \cite{Gill}, pp. 237--241), i.e.
$\,\nabla\times\, ({\bf u},0)+(0,0,l)={\bf 0}$.

\subsection{Small $c_{0}$ expansion}\label{small_c0} The parameter
$c_{0}$ is not singular, therefore the expansion can be bound by a
standard way as series with respect to $c_0$, converging for small
$c_0$ \cite{naife}.

To find the first term  $Q_{0}(t)$ in the expansion of $Q(t)$ we
notice that $Q_{0}(t)$ satisfies a matrix Riccati equation
\begin{equation} \label{Q0}
\dot Q_{0}=-Q_{0}^{2}-lLQ_{0}.
\end{equation}
This equation can be linearized (e.g. \cite{reid}) and solved
explicitly. Indeed, let us assume that $\det Q_{0} \ne 0$ and denote
$U=Q_{0}^{-1}$. This matrix satisfies linear matrix equation
\begin{equation} \label{U}
\dot U=E+lUL.
\end{equation}
It has a general solution
$$
U(t)=\begin{pmatrix}
C_{3}\cos lt+C_{4}\sin lt & C_{3}\cos lt -C_{4}\sin lt-\frac{1}{l} \\
\frac{1}{l}+C_{1}\cos lt +C_{2}\sin lt & C_{2}\cos lt-C_{1}\sin lt
\end{pmatrix}
$$
%$$U(t)=\begin{pmatrix}
%C_{3}\cos lt+C_{4}\sin lt & C_{3}\cos lt -C_{4}\sin lt-\frac{1}{l} \\
%\frac{1}{l}+C_{1}\cos lt +C_{2}\sin lt & C_{2}\cos lt-C_{1}\sin lt
%\end{pmatrix}$$
with $C_1=-\frac{1}{l}+\frac{c(0)}{\det{Q_0(0)}}$,
$C_2=\frac{a(0)}{\det{Q_0(0)}}$,
$C_3=\frac{1}{l}-\frac{b(0)}{\det{Q_0(0)}}$,
$C_4=\frac{d(0)}{\det{Q_0(0)}}$.
% with arbitrary constants
%$C_i,\,i=1,\dots,4.$
Thus,
%$$Q_{0}(t)=\dfrac{l}{1+l^{2}(C_{2}C_{4}-C_{1}C_{3})+(C_{2}+C_{4})l\sin(lt)+(C_{1}
%-C_{3})l\cos(lt)}\times $$
\begin{equation}Q_{0}(t)=\frac{l}{\det U(t)}
\begin{pmatrix}
-C_{2}l\cos lt+C_{1}l\sin lt  & 1-lC_{3}\cos lt +C_{4}l\sin lt \\
-1-lC_{1}\cos lt-lC_{2}\sin lt & lC_{4}\cos lt+lC_{3}\sin lt
\end{pmatrix}, \label{qsol}
\end{equation}
with $\det U(t)= 1+l^{2}(C_{2}C_{4}-C_{1}C_{3})+(C_{2}+C_{4})l\sin
lt +(C_{1} -C_{3})l\cos lt.$

\begin{remark} Analogously to Sec.\eqref{s1}, case $A=0$, system
\eqref{U} can be written as two separate equations for functions
$z_1,\,z_2: {\mathbb C}\mapsto {\mathbb C}$.
\end{remark}

Equation for the first term  $R_{0}(t)$ in the expansion of $R(t)$
is the following:
\begin{equation}\label{R0}
\dot R_{0}+2R_{0}Q_{0}+(\gamma-1)\mbox{\bf tr}Q_{0} R_{0}=0.
\end{equation}
 This is a homogeneous  linear matrix equation with
respect to $R_0$. It can be readily shown that the elements of
symmetric matrix $R_{0}(t)$, $(R_{0})_{ij}(t),$ $i,j=1,2$, can be
found in the form
  %ìîæíî èñêàòü â ñëåäóþùåì âèäå
\begin{equation*}\displaystyle
(R_{0})_{ij}=\frac{1}{(\det
U(t))^{\gamma}}((A_{0})_{ij}+(A_{1})_{ij}\sin(lt)+(A_{2})_{ij}\cos(lt)),\label{rsol} \qquad %\eesol
\end{equation*}
where 9 unknown coefficients $(A_{k})_{ij},$ $i,j=1,2,$ $k=0,1,2;$
$(A_{k})_{ij}=(A_{k})_{ji}$ have to be determined by substitution to
\eqref{R0}.

\begin{proposition}\label{prop1}
System \eqref{Q0} has two equilibria, both are stable in the
Lyapunov sense:
\begin{itemize}
\item 1) $a=d=0$, $b=-c=l$,
\item 2) $a=b=c=d=0$.
\end{itemize}
They are separated by unstable manyfold
\begin{equation}\label{im}
b=\frac{l}{2},\quad c=-\frac{l}{2},\quad a=d.
\end{equation}
\end{proposition}

{\sc Proof.} The functions $$V(a,b,c,d)= \frac{ a^{2}+ d^{2}+
 \left( b-{\frac {ad-bc}{l}} \right) ^{2}+ \left( c+{\frac {ad-bc}{l}} \right) ^{2} }{( ad-bc)^{2}}
 $$
can be considered as Lyapunov functions for equilibria 1). It can be
readily computed that $\frac{d}{dt} V (a,b,c,d)=0$  on the solutions
of system \eqref{Q0}. Thus, the equilibrium is stable non
asymptotically and in the vicinity of equilibria every trajectory is
closed. The change of variables $a_1=a$, $b_1=b-l$, $c_1=c+l$,
$d_1=d$ maps the origin into $(0, -l, l, 0)$ and the existence of
the Lyapunov function
$$V(a_1,b_1,c_1,d_1)= \frac{ a_1^{2}+ d_1^{2}+
 \left( b_1+{\frac {a_1d_1-b_1c_1}{l}} \right) ^{2}+ \left( c_1-{\frac {a_1d_1-b_1c_1}{l}} \right) ^{2} }
 {( a_1d_1-b_1c_1)^{2}}
 $$
 proves the stability of the second equilibrium.

 Further, it is easy to check that if  \eqref{im} holds, then
 the system can be reduced to one equation $\dot a =-a^2-l^2/4$. Therefore
 $a\to -\infty$, $d\to -\infty$ for any initial data satisfying
 \eqref{im}.
$\square $

 \begin{remark} \label{rem1} The equilibria of system \eqref{Q0}, \eqref{R0} correspond to
 the equilibria \eqref{equililria} of full system \eqref{m1},
 \eqref{m2}  only if $A^*=0$. The stability of equilibria of the
zero approximation system does not imply stability of the
equilibrium of full system. Moreover, for the full system these two
equilibrium and unstable manyfold glue together. Thus, $c_0$ is a
parameter of bifurcation for the system \eqref{m1}, \eqref{m2}.
\end{remark}

\subsection{Range of instability}

\begin{proposition}\label{TStable}
%{ \rm  (see also \cite{RYH2015})}
If
$$b^*<
\frac{1-\sqrt{2}}{2}\,l \quad \mbox{or} \quad
b^*>\frac{1+\sqrt{2}}{2}\,l>l,$$ then the equilibrium of system
\eqref{m1}, \eqref{m2} is unstable.
\end{proposition}

{\em Proof}.
  The
eigenvalues of matrix corresponding to the linearization at the
equilibrium point of the system \eqref{m1}, \eqref{m2},
\eqref{first_int} are the following:
$$\lambda_{1,2}=\pm
\sqrt{-(2(2-\gamma)b^*(b^*-l)+l^2)},$$  $$ \lambda_{3,4}=\pm\sqrt{
2}\,\sqrt{-l\left(b^*+\frac{l}{4}\right)+
\sqrt{\left(b^*+\frac{l}{2}\right)^2\left(\frac{l^2}{4}+b^*
l-(b^*)^2\right)}},$$
$$
\lambda_{5,6}=\pm\sqrt{ 2}\,\sqrt{-l\left(b^*+\frac{l}{4}\right)-
\sqrt{\left(b^*+\frac{l}{2}\right)^2\left(\frac{l^2}{4}+b^*
l-(b^*)^2\right)}}.$$

 Since $(2-\gamma)b^*(b^*-l)+l^2>0$ for
$\gamma\in(1,2)$, then $\Re(\lambda_{1,2})=0$. Eigenvalues
$\lambda_i,\,i=3,4,5,6$ have zero real part if and only if $b^*$
satisfies the following inequalities simultaneously: $
l(b^*+\frac{l}{4})\ge 0,\quad \frac{l^2}{4}+b^* l-(b^*)^2>0,$ $
l^2\left(b^*+\frac{l}{4}\right)^2>\left(b^*+\frac{l}{2}\right)^2\left(\frac{l^2}{4}+b^*
l-(b^*)^2\right),$ that is $b^*\in
\big[\frac{1-\sqrt{2}}{2}\,l,\,\frac{1+\sqrt{2}}{2}\,l \big]$. For
others values of $b^*$ the eigenvalues $\lambda_{3,4,5,6}=\pm \alpha
\pm i \beta, \,\alpha\ne 0, \beta \ne 0$, therefore there exist an
eigenvalue with a positive real part. Thus, the Lyapunov theorem
implies instability of the equilibrium for $b^*<
\frac{1-\sqrt{2}}{2}\,l$ and $b^*>\frac{1+\sqrt{2}}{2}\,l>l.$
$\square$
%\end{proof}

\begin{remark} If the coordinate system is not rotating ($l=0$), then the
vortex is always cyclonic (the anticyclonic domain shrinks as $l\to
0$). The equilibrium point is  unstable in the Lyapunov sense both
in axisymmertic and general case. Nevertheless,  in the axisymmetric
case the equilibrium has a type of stable/unstable node and is
quasi-asymptotically stable, whereas in general case the matrix of
linearization has eigenvalues with nonzero real parts. Thus, we can
see that the rotation has a stabilizing effect.
\end{remark}

\subsection{Range of possible stability}

If $b^*\in  \Sigma=\Sigma_+$, $\Sigma_+=
\left(\frac{1-\sqrt{2}}{2}l,0\right)\cup
\left(l,\frac{1+\sqrt{2}}{2}l\right)$ (cyclonic case) or $b^*\in
\Sigma_-$, $\Sigma_-= \left(0,l\right)$ (anticyclonic case),  then
the matrix, corresponding to the system, linearized at the
equilibrium, has 3 pairs of pure imaginary complex conjugate roots
which can be written as $\pm i\,\omega_j \,,$  $j=1, 2, 3$,
$\omega_j\in \mathbb R$.  For the range of parameters under
consideration ($\gamma\in (1,2)$) the roots are simple, for
$\gamma>3+\sqrt{2}$ the roots can be multiple.

In the boundary cases $b^*=\frac{1-\sqrt{2}}{2}l$ and
$b^*=\frac{1+\sqrt{2}}{2}l$ one of the pairs of pure imaginary
complex conjugate roots is multiple.

 The degenerate cases $b^*=0$ and $b^*=l$, where $A^*=0$
were considered in Sec.\eqref{small_c0}, see Remark \eqref{rem1}.

\subsection{Semi-analytical method to prove stability: non-resonant
frequencies}

We are going  to prove that in the general case of rationally
independent frequencies almost all trajectories in
$\epsilon$-neighborhood of the equilibrium are quasi-periodic. This
means that the equilibrium is "practically" stable in the Lyapunov
sense. We apply a semi-analytical method based on the Bibikov
theorem \cite{Bibikov} (Theorem 15.5).

Let us consider system
\begin{eqnarray}\label{Asyst}
\dot X = {\mathcal A} X + {\mathcal P}(X),\quad  X\in {\mathbb R}^n,
\end{eqnarray}
where a constant matrix $A$ has purely imaginary eigenvalues $\pm i
\omega_j $, $j=1,\dots, m$, $n=2m$, the frequencies $\omega_j$ are
rationally independent. The vector-valued function ${\mathcal
P}(X)$, which does not contain free and linear terms. The system can
be written in diagonalized form
\begin{eqnarray}\label{diagon_form}
 \dot y_k =& i\omega_k y_k + Y_k(y,\bar y),\\
 \dot {\bar y}_k =& -i\omega_k \bar y_k + \bar Y_k(y,\bar
 y),\nonumber
\end{eqnarray}
Further, \eqref{diagon_form} is formally equivalent to its  normal
form
\begin{eqnarray}\label{normal_form}
 \dot y_k =& y_k (i\omega_k y_k + P_k(y,\bar y)),\\
 \dot {\bar y}_k =& \bar y_k (-i\omega_k \bar y_k + \bar P_k(y,\bar
 y)),\nonumber
\end{eqnarray}
where $P_k( y \bar y)$ denotes a (formal) series in powers of
products $y_1 \bar y_1,\dots , y_m \bar y_m$  without constant terms
\cite{Bruno}. The normalizing transform has the form
\begin{eqnarray}\label{normal_trans}
x_k=y_k+h_k(y_k, \bar y_k), \quad \bar x_k=\bar y_k+\bar h_k(y_k,
\bar y_k),
\end{eqnarray}
where the series $h_k(y_k, \bar y_k)$ are also formal.

 Further, let us assume that the neutrality
condition takes place:
\begin{eqnarray}\label{H-condition}
 P_k (y_1 \bar y_1, \dots, y_m \bar y_m) = i H_k(y_1 \bar y_1, \dots, y_m \bar y_m ),
\end{eqnarray}
where $H_k$ are series with real coefficients. Then system
\eqref{normal_form} has as integral surfaces  invariant
$m$-dimensional tori $y_k \bar y_k = c_k
> 0,$ $ k = 1, ..., m$, and possess quasi-periodic solutions.
If the equivalence of systems \eqref{Asyst} and \eqref{normal_form}
was not only formal, but analytical, then the invariant tori to
system \eqref{normal_form}  would correspond to invariant tori to
system \eqref{Asyst}.

Theorem 15.1 \cite{Bibikov} implies that despite of divergence of
normalizing transform \eqref{normal_trans}, in some sense "most"\,
of invariant tori to the system \eqref{normal_form} correspond to
invariant tori to system \eqref{Asyst}. Namely, there exist
$\epsilon
> 0$ and series $h_k (y, \bar y , \rho)$, $\rho=y \bar y$,
$k=l,...,n$, converge in $\epsilon$ - neighborhood of the origin
$\mathcal V_\epsilon$ for every $\rho$ belonging to a measurable set
$\mathcal M_\epsilon\in \mathcal V_\epsilon $, where
$\lim\limits_{\epsilon\to 0}\frac{{\rm mes}\,\mathcal
M_\epsilon}{{\rm mes}\,\mathcal V_\epsilon}=1$, such that change of
variables \eqref{normal_trans} reduces system \eqref{Asyst}  to
\eqref{normal_form}, where $H_k$ are convergent for $y\in {\rm
mes}\,\mathcal V_\epsilon$, for every $\rho \in {\rm mes}\,\mathcal
M_\epsilon$ and have real coefficients.

To apply the Bibikov theorem we have  to reduce the system
\eqref{m1}, \eqref{m2}, \eqref{first_int} to normal form.

Namely, we have to realize the following algorithm.
%\begin{enumerate}
\begin{itemize}
\item The system \eqref{m1}, \eqref{m2}, \eqref{first_int} has to be written in the form \eqref{diagon_form} at
the equilibrium point $a=d=0$, $b=-c=b^*$,
$A=C=A^{*}=\frac{b^*(b^*-l)}{2c_0}$, $B=0$:
    \begin{equation} \label{form}
    \frac{dx_{\nu}}{dt}=\lambda_{\nu} x_{\nu} + \sum a_{jh}^{\nu}x_{j}x_{h}+\sum b_{jhk}^{\nu}x_{j}x_{h}x_{k}+...,
    \end{equation}
    where indices $\nu$ take values $\pm 1, \pm 2, \pm 3$;
    $\bar{\lambda}_{\nu}=\lambda_{-\nu}$ and
    $\bar{x}_{\nu}=x_{-\nu}$.
    Here
    $x_{\nu}$ are new variables,  $\lambda_{\nu}$ and
    $\lambda_{-\nu}$ correspond to the complex conjugate
    eigenvalues, coefficients $ a_{jh}^{\nu}$ and $ b_{jhk}^{\nu}$
    are complex-valued and symmetrized, $j, h, k, \nu =\mp 1, \mp 2, \mp
    3$.

    \item According to the Bruno theorem  (\cite{Bruno},\cite{Bruno1}
    \cite{Starzhiskii}, \cite{Bibikov}) there exists a formal change of variables
     $$x_{j}=y_{j}+\sum
    \alpha_{lm}^{j}y_{l}y_{m}+\sum
    \beta_{lmn}^{j}y_{l}y_{m}y_{n}+...,$$
 where  $ \alpha_{lm}^{j} = \alpha_{ml}^{j}$, $  \beta_{lmn}^{j}= {\rm id},$ $j,l,m,n=\mp 1, \mp 2,
 \mp 3$, reducing the system
    \eqref{form} to normal form:
    \begin{equation}\label{n_f}
    \frac{dy_{\nu}}{dt}=\lambda_{\nu} y_{\nu}+y_{\nu} S(\,{y},\bar y \,)=
    \lambda_{\nu} y_{\nu}+y_{\nu}\sum\limits_{({\bf \Lambda, Q})=0}g_{\nu {\bf Q}}
    y_{1}^{q_1}
    y_{-1}^{q_{-1}}
    y_{2}^{q_2}
    y_{-2}^{q_{-2}}
    y_{3}^{q_3}
    y_{-3}^{q_{-3}},
 \end{equation}
 where $\nu=\mp1, \mp 2, \mp 3$, $q_j\in \mathbb Z$, $q_\nu\ge -1$,
 $q_j >0, \,j\ne \nu $, $\sum q_n \ge 1$. For our case condition $({\bf \Lambda,
 Q})=0$ means
\begin{equation*}
\omega_1(q_1-q_{-1})+\omega_2(q_2-q_{-2})+\omega_3(q_3-q_{-3})=0,
\quad \omega_\nu= \Im \lambda_\nu.
 \end{equation*}
 If we restrict ourselves by the non-resonant case, where $\omega_\nu$ are rationally independent,
  we obtain $q_\nu = q_{-\nu},\, \nu=1, 2, 3.$
  Thus, the series $S(\,{y},\bar y)$ contains infinitely many terms.

\item According to the Bibikov theorem to prove that almost all trajectories in
$\epsilon$-neighborhood of the equilibrium
 $a=d=0$, $b=-c=b^*$,
$A=C=A^{*}=\frac{b^*(b^*-l)}{2c_0}$, $B=0$ are quasi-periodic we
have to show that
$S(\,\vec{y}\,)=iH(y_{1}y_{-1},y_{2}y_{-2},y_{3}y_{-3})$, where $H$
is a real valued vector-function. Thus, we have to check that the
coefficients $g_{\nu {\bf Q}}$
are pure imaginary. \\
%\end{enumerate}
\end{itemize}

All the steps are standard, however, the computations are very
cumbersome, therefore, we are  performing them by means of computer
algebra packet.

\subsection{Method of computing $g_{\nu {\bf Q}}$, truncated case.}

In \cite{Starzhiskii}, Ch.VIII, Sec.4, the author analyzes exists
resonances and normal forms of analytic autonomous (not necessarily
conservative) sixth-order systems with three pairs of distinct pure
imaginary eigenvalues of the matrix of the linear part, as in our
case. Provided the normal form  \eqref{n_f} is truncated to terms of
power not higher than three, there exist  explicit formulae for
calculation of coefficients of normalizing transformation and normal
forms. %Moreover, he analyzes all possible resonances for this case.

Namely, the truncated normal form \eqref{n_f}  is
\begin{equation}\label{n_f_3}
    \frac{dy_{\nu}}{dt}=\lambda_{\nu} y_{\nu}+y_{\nu} S_3(\,{y},\bar y \,)=
\lambda_{\nu} y_{\nu}+y_{\nu}(g_{1}^{\nu}y_{1}y_{-1}+
    g_{2}^{\nu}y_{2}y_{-2}+g_{3}^{\nu}y_{3}y_{-3}).
    \end{equation}

Method of computing  $g_{i}^{j}$  is the following
\cite{Starzhiskii}.
\begin{enumerate}
\item We denote the vector of solutions $(A,B,C,a,c,d)^{T}$ as $Z$ and re-write the initial system
 \eqref{m1}, \eqref{m2}, \eqref{first_int} as
    \begin{equation}\label{equat}
    Z'=GZ+F(Z),
    \end{equation}
    where $G$ is a matrix of linearization at the equlibrium, $F(Z)$ is the nonlinear part.
\item We reduce   $G$ to its diagonalized form $D$ by means of the non-degenerate matrix  $C$, such that
  $G=C^{-1}DC$. The change of variables $Y=CZ$
    reduces  \eqref{equat} to
    $$Y'=DY+CF(C^{-1}Y).$$
\item We expand the matrix  $CF(C^{-1}Y)$ of nonlinear part to the Taylor series in the new variables
at the equilibrium  and find the coefficients of second order,
$a_{jh}^{\nu}$, and third order, $b_{jhk}^{\nu}$.
\item Further, we find coefficients $\alpha_{lm}^{\nu},g_{h}^{\nu}$
by the following formulae:
\begin{equation*}
    \alpha_{lm}^{\nu}=\frac{a_{lm}}{\lambda_{l}+\lambda_{m}-\lambda_{\nu}};\label{alpha}
    \end{equation*}
    $$g_{|\nu|}^{\nu}=3 b_{\nu \nu -\nu}^{\nu}+2\sum_{j} (2a_{\nu j}^{\nu}\alpha_{\nu -\nu}^{j}+a_{-\nu j}^{\nu}\alpha_{\nu \nu}^{j});$$
    $$g_{\nu}^{h}=
    6 b_{\nu h -h}^{\nu}+4\sum_{j} (a_{\nu j}^{\nu}\alpha_{h -h}^{j}+
    a_{hj}^{\nu}\alpha_{-h \nu}^{j}+a_{-h j}^{\nu}\alpha_{\nu h}^{j}),$$
    where $h \neq |\nu|; \nu=\pm 1, \pm 2, \pm 3$.
\end{enumerate}

We performed computations with the step $0.01$ with respect to
$\gamma$ and $b^{*}$, and they confirms that the real  parts of
$g_{\nu}^{h}$ are zero with a good reliability inside of $\Sigma_+$.
The real parts of $g_{\nu}^{h}$ do not vanish in a very small
neighborhood  of  boundary points $ \frac{1-\sqrt{2}}{2}l$ and
$\frac{1+\sqrt{2}}{2}l$ and points $0$ and $1$, where the
anticyclonic domain changes to %anomalous low
cyclonic one. Therefore,  the necessary condition for the stability
does not hold for the full normal form \eqref{n_f} too, and the
equilibrium is instable there. Thus, we check numerically that
except of small neighborhood of these points the neutrality
condition \eqref{H-condition} holds for the truncated up to third
terms normal form \eqref{n_f_3}.

 %Ìíèìûå ÷àñòè êîýôôèöèåíòîâ $g_{i}^{j}$ ïðèâåäåíû íà {\bf ðèñ. 2-5}.

\subsection{Resonant frequencies}

It is interesting that all resonances up to order 3 are very close
to the boundaries of $\Sigma$. Indeed, if the frequencies
$\omega_j$, $j=1,2,3$ are ordered as
  $0<\omega_1<\omega_2<\omega_3$, then the resonance values of $b^*$
  can be found from equalities
$$
\omega_1=\frac{\omega_3}{2},\frac{\omega_3}{3}; \quad
\omega_2=\frac{\omega_3}{2}, \frac{\omega_3}{3}; \quad
\frac{\omega_1}{\omega_2}=\frac{\omega_3}{2},\frac{\omega_3}{3};
$$
$$
\omega_1+\omega_2=\omega_3; \quad 2\omega_1+\omega_2=\omega_3; \quad
2\omega_2\pm \omega_1 =\omega_3.
$$
It is easy to check that if $\frac{b^*}{l}\in
    \left(1,\frac{1+\sqrt{2}}{2}\right)$, then
$\omega_1=\Im \lambda_1$, $\omega_2=\Im \lambda_5$, $\omega_3=\Im
\lambda_3$.

 If $\frac{b^*}{l}\in
    \left(\frac{1-\sqrt{2}}{2},0\right)$, then
$\omega_1=\Im \lambda_5$, $\omega_2=\Im \lambda_3$, $\omega_3=\Im
\lambda_3.$ The computations made for  $\gamma\in (1,2)$ show, that
all resonances $b^*/l$ in the domain $\Sigma_+$ are close to the
boundaries $\frac{1-\sqrt{2}}{2} \approx -0.2071$ and
$\frac{1+\sqrt{2}}{2} = 1.2071$. Moreover, a lot of resonances
accumulate near the point $\frac{1-\sqrt{2}}{2}$.

The resonances of order $n$ between frequencies $\omega_2=|\Im
\lambda_i|$, $i=3,4$ and $\omega_3=|\Im \lambda_j|$, $j=5,6$ have
place for $\frac{b^*_n}{l}=\frac{n\pm \sqrt{n/2}(n+1)}{n^2+1}\cap
\overline{\Sigma_+\cup \Sigma_-}$. These values do not depend of
$\gamma$. As follows from this formula, $\frac{b^*_n}{l}\to 0$ as
$n\to \infty$, all positive $b^*_n$ are in the anticyclonic domain,
negative $b^*_n$ fall in the cyclonic domain starting from $n=22$.

The resonances of order $n$ between frequencies $\omega_2=|\Im
\lambda_j|$, $j=3,4$ and $\omega_1=|\Im \lambda_k|$, $k=1,2$ also
have places in $\overline{\Sigma_+\cup \Sigma_-}$. For example, if
$\gamma=2$, then an explicit  formula can be obtained:
$\frac{b^*_n}{l}=\frac{1\pm \sqrt{2n-1}(n+1)}{2n}\to 0$ as $n\to
\infty$. The computations show that  in the cyclonic domain
$\Sigma_+$ for all $\gamma \in(1,2)$ at least first resonances are
very close to the left boundary of the domain, $\frac{1-\sqrt{2}}{2}
.$

%Thus, summarizing the reasoning, we can notice that most "dangerous"
%resonance case fall in the anticyclonic domain $\Sigma_-$.

\subsection{Conclusion on the nonlinear stability}

Let us summarize our reasoning.

\begin{quote}{\em Inside  $\Sigma=\big[ \frac{1-\sqrt{2}}{2},
\frac{1+\sqrt{2}}{2}\big]$ there exist domains $\Sigma_{instable}$,
containing points, corresponding to the values of parameter $b^*/l$,
where the equilibrium \eqref{equililriaf} is instable. In the domain
$\Sigma\backslash\Sigma_{instable}$, except may be a discrete number
of points, corresponding to the resonance frequencies, the
sufficient condition for  stability of the equilibrium holds
"numerically"
 up to the term of third order. }
\end{quote}

Of course, we cannot call this statement  a theorem, since
 the "numerical proof" is the computation in a finite
number of points, how frequently they would be located.

 In fact, our computations
only give an argument in favor of stability on $\Sigma_{stable}$.

\begin{figure}[h]
%\begin{minipage}{0.5\columnwidth}
\centerline{\includegraphics[width=0.5\columnwidth]{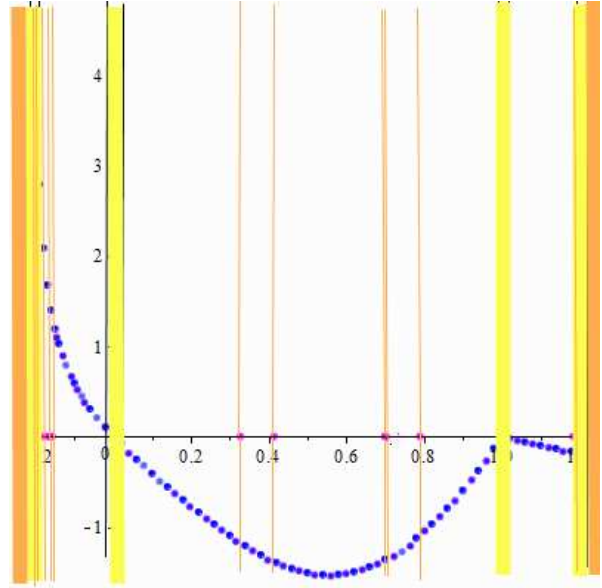}}
\caption{Imaginary part of coefficients $g_3^3$ (dots) on $\Sigma$,
resonances of the third order (vertical lines), domain
$\Sigma_{instable} $, where the imaginary parts of several
coefficients $g_i^j$ do not vanish (yellow vertical lines) for
$\gamma=9/7$. }\label{pic9-7n.ps}
\end{figure}

Fig.\ref{pic9-7n.ps} presents the positions of resonances up to
third order (vertical lines) and the domain containing points, where
the real part of coefficients $g_{i}^{j}$ does not vanish for
$\gamma=9/7$ (this value of $\gamma$ corresponds to the 2D model of
atmosphere, averaged over the height \cite{Obukhov}).

Direct numerical computation from the system \eqref{m1}, \eqref{m2},
confirm our hypothesis  on stability of equilibrium on
$\Sigma_{stable}$.

%The cases of possible resonant frequencies correspond to several
%values of $b^*$, very close to the boundaries of $\Sigma$.

Of course, our results concern stability with respect to small
perturbations and the basin of attraction can be very small. The
initial data that fall into the basin of attraction of a stable
equilibrium can be found only numerically except of the case $b^*=0$
and $b^*=l$, where we obtained exact solution in Sec.\ref{small_c0}.
However, even in this case the domain in the four-dimensional space
of parameters  $a_0=a(0), b_0=b(0), c_0=c(0), d_0=d(0)$ is very
complicated. Indeed, as follows from \eqref{qsol}, the trajectory
will be periodic and therefore does not move away from the
equilibrium for such initial data that $\det U(t)= 1+l^2(C_2 C_4-C_1
C_3)+(C_2+C_4)l\sin lt +(C_1 -C_3)l\cos lt\ne 0$ for all $t>0$.
Thus, the initial data corresponding to the basin of attraction have
to satisfy the following condition:
\begin{equation*}- (a_0 d_0+b_0 c_0)^2 l^2+2\,
 ( b_0+c_0)  (  a^2_0 d^2_0- b^2_0 c^2_0)   l+
 (  ( a_0- d_0)^2-4 b_0 c_0)  (a_0 d_0-b_0 c_0)^2<0.
\end{equation*} In particular, this condition implies that increasing
of $l$ at fixed parameters $a_0, b_0, c_0, d_0$ stabilizes the
motion.

The case $l=0$ ie exceptional. As follows from the results of
Sec.\ref{s1}, in this case there is no nontrivial steady state
solution an even the trivial equilibrium is unstable in the Lyapunov
sense. Nevertheless, the oscillating regime can arise from elliptic
perturbation of the axisymmetric state. Namely, as it was mentioned,
in \cite{Anisimov_Lysikov} it was shown that the system
\eqref{mathcalF}, an analog of \eqref{m1}, \eqref{m2} for the case
of the finite mass, is integrable for $\gamma=2$. This value of
$\gamma$ corresponds to the shallow water equations. The formulae
are very complicated and contain elliptic integrals, therefore it is
hard to analyze them. Nevertheless, in \cite{Bogoyavlensky}, Ch.7,
Sec.3 it is shown qualitatively that in this problem there exists
oscillations and their period is estimated.

\section{Comparison with numerics: localized vortex}\label{Sec3}

We studied stability of the vortex corresponding to a very specific
class of solutions. In fact, real vortices have the linear profile
of velocity only near its center (see the discussion in
\cite{RYH2010}), therefore  our results on the nonlinear stability
or instability  for  the "toy" solution \eqref{u-form},
\eqref{pi-form} not obliged to hold for "realistic" solutions.

According \cite{RYH2012}, \cite{R15}  the steady solution to
\eqref{main}, \eqref{pi} can be constructed as follows. Let us
consider arbitrary smooth enough stream function $\Phi=\Phi \left(
{x_1}^{2}+{x_2}^{2} \right)$. Then
\begin{equation}\label{U_formula}
{\bf U}=\nabla_\bot \Phi = (\Phi_{x_2}, - \Phi_{x_1})
\end{equation}
and
\begin{equation}\label{pi_formula}
\Pi=-  \frac{1}{c_0}\,\left[2 l\Phi + \int(\Phi_{x_2} \Phi_{x_1
x_2}-\Phi_{x_1} \Phi_{x_2 x_2})\,dx_1 + \int(-\Phi_{x_2} \Phi_{x_1
x_1}+\Phi_{x_1} \Phi_{x_1 x_2})\,dx_2 \right].
\end{equation}

In \cite{RYH2012} we choose the  solution of such class with
exponential decay for numerical computations in the context of big
atmospherical vortices such as typhoons,i.e. this solution has the
form
\begin{equation}\label{ic_u} u_1^0\,=\,k\,B_0\,\sigma\,x_2\,e^{-\frac{\sigma}{2}\, \left(
{x_1}^{2}+{x_2}^{2} \right)},\qquad u_2^0\,=\,
-B_0\,\sigma\,x_1\,e^{-\frac{\sigma}{2}\, \left( {x_1}^{2}+{x_2}^{2}
\right)},\quad k=1,
\end{equation}
\begin{equation}\label{ic_pi}
\Pi^0\,=\,-\,\frac{1}{2c_0}\,\left(\,B_0^2\,\sigma\,e^{-\sigma\,(x_1^2+x_2^2)}\,
-2\,l\,B_0\,e^{-\frac{\sigma}{2}\,(x_1^2+x_2^2)}\right)+R_0.
\end{equation}
is case corresponds to $\Phi=-B_0\,e^{-\frac{\sigma}{2}\, \left(
{x_1}^{2}+{x_2}^{2} \right)}$.

Nevertheless, based on experimental data, the meteorologists use a
piecewise-continuous profile of tangential wind speed $V(r)$  such
as $V(r)={\rm const}\cdot r $ for $r<R$ and $V(r)= {\rm const}\cdot
r^{-\frac{5}{8}}$ for $r>R$, where $R$ is the radius of the cyclone
eye \cite{vanderman}, so we study the numerical solution for initial
conditions with "natural" slower decay rate such as $x_1^2+x_2^2\to
\infty$\footnote{See also \cite{Kloosterziel_1} in the context of
the decay of vortex at infinity.}. To construct the steady solution,
we use the following stream function
$$\Phi=\frac{B_0}{q}\left(1+\frac{\sigma}{2}\, \left( {x_1}^{2}+{x_2}^{2}
\right)\right)^{-q},\quad q>0,$$ which gives a family of initial
data :
\begin{eqnarray}\label{ic_uq}
u_1^0\,=\,k\,B_0\,\sigma\,x_2\,\left(1+\frac{\sigma}{2}\, \left(
{x_1}^{2}+{x_2}^{2} \right)\right)^{-q-1},\\ u_2^0\,=\,
-\,B_0\,\sigma\,x_1\,\left(1+\frac{\sigma}{2}\, \left(
{x_1}^{2}+{x_2}^{2} \right)\right)^{-q-1},\nonumber
\end{eqnarray}
\begin{equation}\label{ic_piq}
\Pi^0\,=\,-\,\frac{2}{c_0}\left[-\frac{B_0 l}{q
}\,\left(1+\frac{\sigma}{2}\, \left( {x_1}^{2}+{x_2}^{2}
\right)\right)^{-q}+ \frac{B_0^2\sigma}{2q+1}\,
\left(1+\frac{\sigma}{2}\, \left( {x_1}^{2}+{x_2}^{2}
\right)\right)^{-2q-1}\right]+R_0,
\end{equation}
They provide the steady solution for $k=1$.

It is easy to check that both velocities \eqref{ic_u} and
\eqref{ic_uq} have a linear profile with $b^*= B_0 \sigma$ for those
solutions close to the origin .

%\begin{remark}
Easy computations show that for the both classes of data
\eqref{ic_u}, \eqref{ic_pi} and \eqref{ic_uq}, \eqref{ic_piq}, the
pressure has maximum or minimum at infinity for the same values of
parameters as for the case of the velocity with linear profile. That
is, \eqref{ic_pi} and \eqref{ic_piq} have a maximum at the origin
provided $B_0 \sigma/l=b^*/l\in (0, 1)$. Nevertheless,  if $B_0
\sigma/l=b^*/l\in (1, s_q)$, $s_q=\frac{2q+1}{q}>1,$ $q>0$,
 the value of $\Pi^0-R_0$ in \eqref{ic_piq} at zero is positive. This means
 that a small domain of low pressure
 exists  on a background of high pressure.   If $q\to \infty$, then $s_q\to 2$.
The data with exponential decay \eqref{ic_u}, \eqref{ic_pi}
correspond to a limit case, and the phenomenon takes place for $B_0
\sigma/l=b^*/l\in (1, 2)$. Thus, we can see that a stronger decay of
initial data at infinity in some sense enlarges an "anomalous low"
domain.
%\end{remark}

Let us analyze whether the range of parameter $ B_0 \sigma$ that
guarantees  stability of the localized vortex like \eqref{ic_u},
\eqref{ic_pi} (or \eqref{ic_uq}, \eqref{ic_piq}) can correspond to
the respective range of stability of parameter $b^*$.

Basically, the answer is negative. Indeed, we have to compute the
value of $G_{x_1}(0)$ (see Sec.\ref{finite_mass}) with $$\rho=
\frac{\gamma}{c_0 (\gamma-1)}(\Pi^0-R_0)^{1/(\gamma-1)}$$ and
estimate how far $G_{x_1}(0)$ is from the value of $A^*$ (see
\eqref{equililriaf}). This requires the computation of improper
integral  with respect to $x_1$ and $x_2$. The integral can be taken
analytically in the exponential case \eqref{ic_pi} with $\gamma=2$.
 It can be readily checked that here
$$G_{x_1}(0)=G_{x_2}(0)=-\frac{1}{2} \frac{\pi(B_0\sigma-8l)B_0}{\sigma^2
c_0^2}, \quad G_{x_1 x_2}(0)=0,$$  $G_{x_1}(0)=G_{x_2}(0)=A^*$ for
$B_0=0$ and $B_0=l\frac{c_0\sigma^3+8\pi}{c_0\sigma^3+\pi}.$ Thus,
for sufficiently small values of $B_0$, the values of $A^*$ and
$G_{x_1}(0)$ are close for any $\sigma$ and $c_0$. For large $B_0$
$A^*$ and $G_{x_1}(0)$, their values are close only for some
specific parameters. This analysis is approximate, since in the
velocity profile is not linear for the localized vortex, considered
in this section, nevertheless, it helps us to understand the
phenomenon.

\begin{figure}[h]
\begin{minipage}{0.33\columnwidth}
\centerline{\includegraphics[width=1\columnwidth]{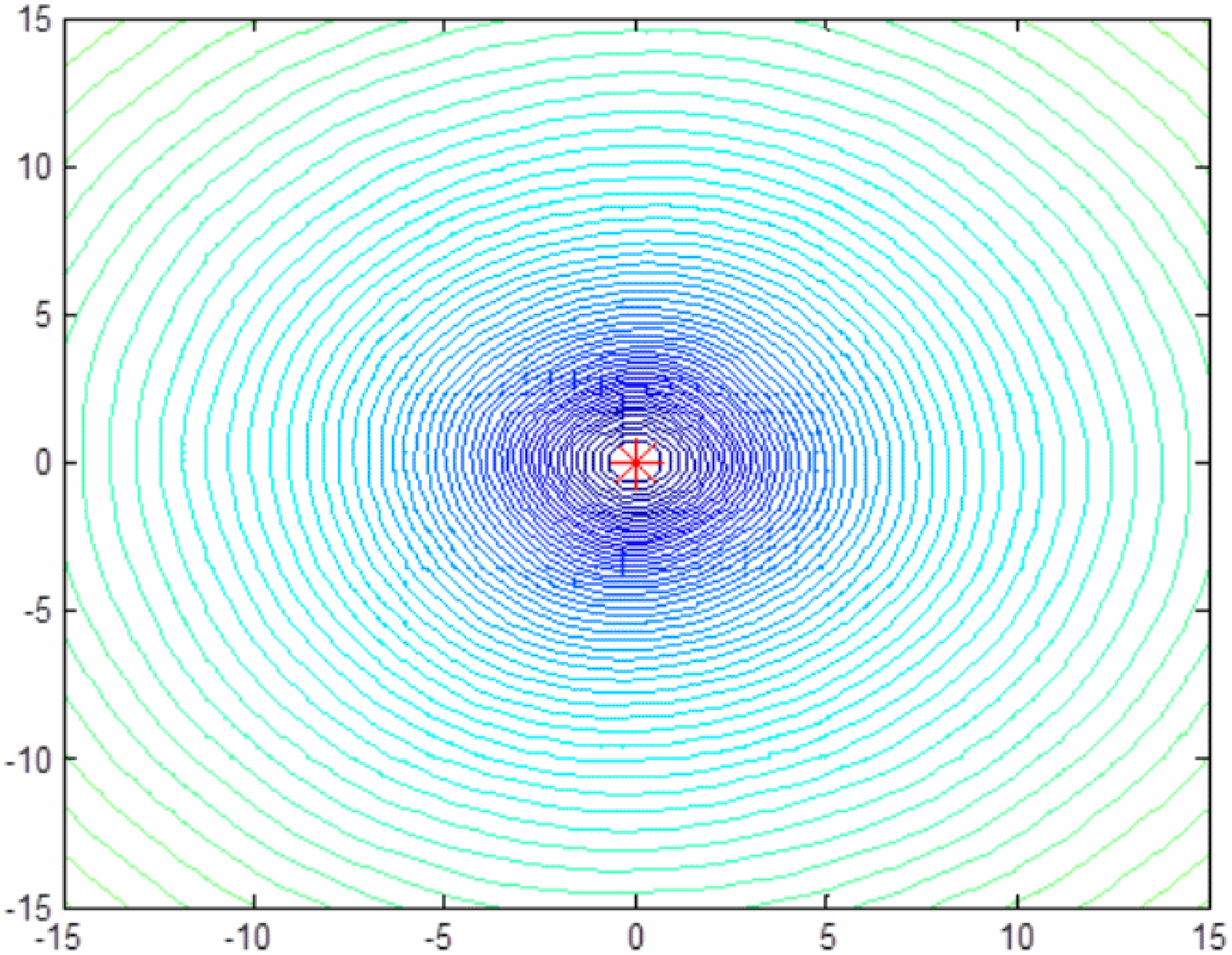}}
%\caption{(a)}
\end{minipage}
%\end{figure}%
%\begin{figure}[h]
\begin{minipage}{0.33\columnwidth}
\centerline{\includegraphics[width=1\columnwidth]{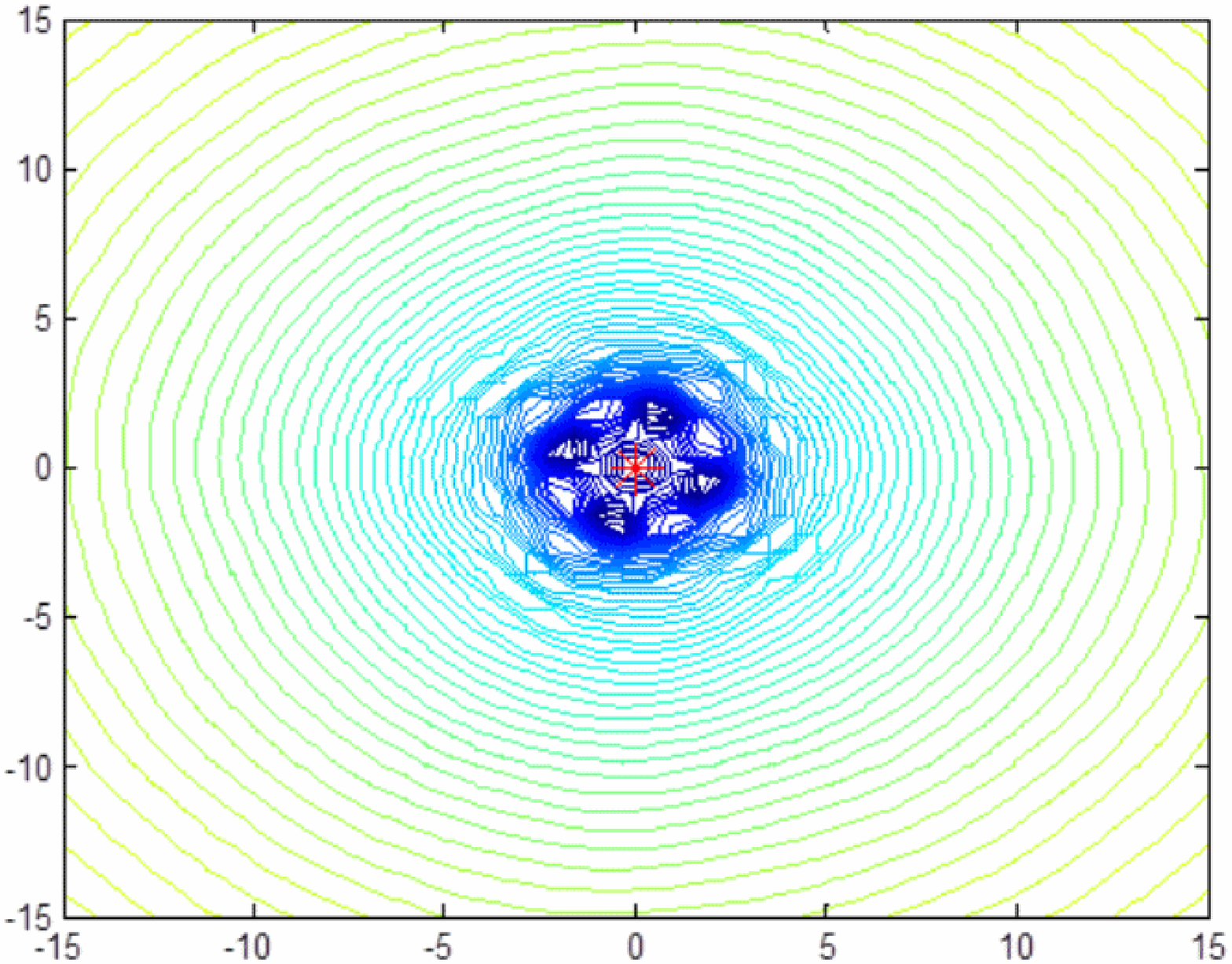}}
%\caption{(b)}
\end{minipage}%
\begin{minipage}{0.33\columnwidth}
\centerline{\includegraphics[width=1\columnwidth]{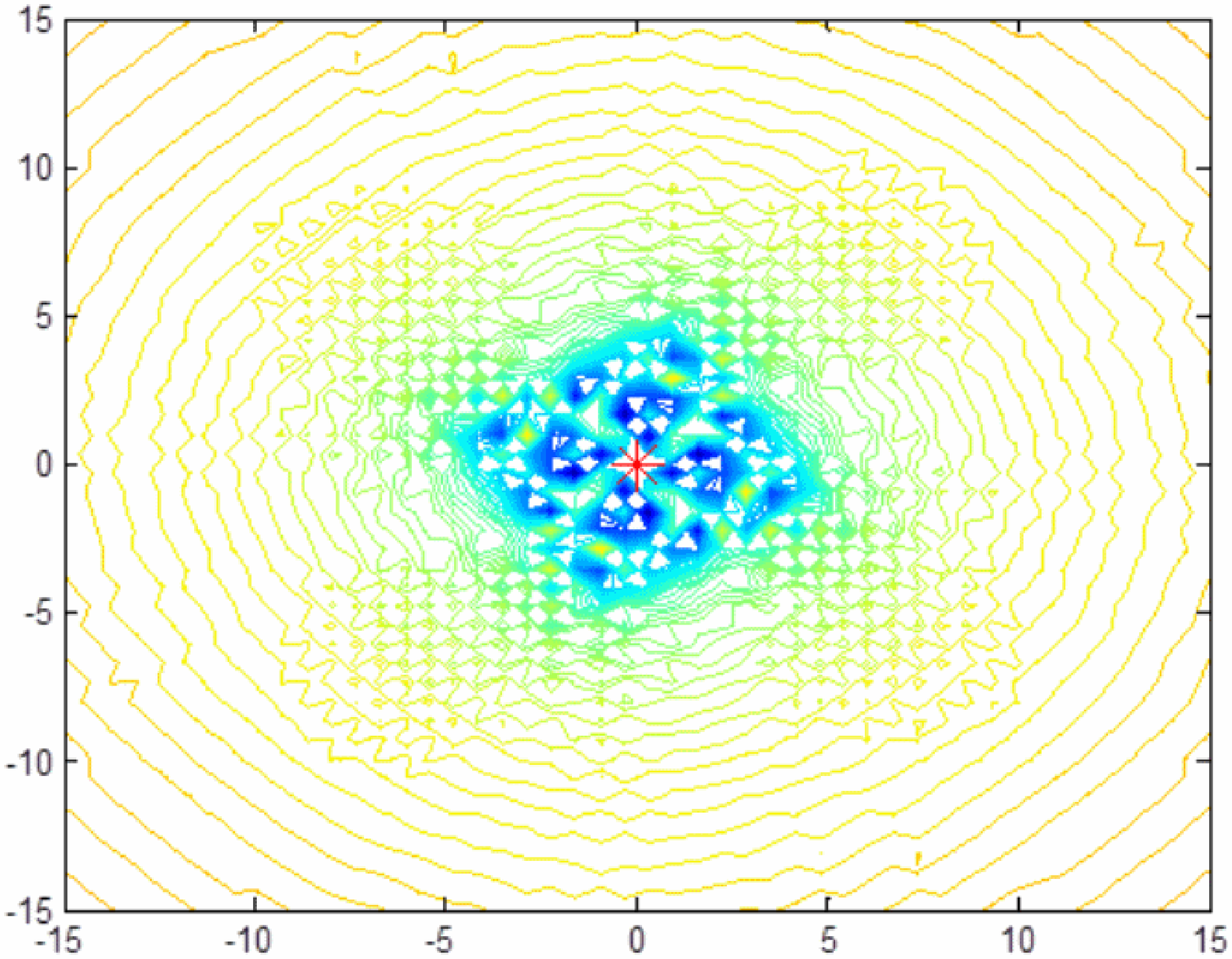}}
%\caption{(b)}
\end{minipage}%
\caption{Transformation of vortex for cyclonic vorticity: $B_0=
-10^{3}$, $B_0=-5\cdot 10^{3}$ and $B_0=-8\cdot 10^{3}$. $k=1.5$,
 $T\approx 24 \rm h$,  the horizontal scale is in kilometers}\label{instability1}
\end{figure}

\begin{figure}[h]
\begin{minipage}{0.33\columnwidth}
\centerline{\includegraphics[width=1\columnwidth]{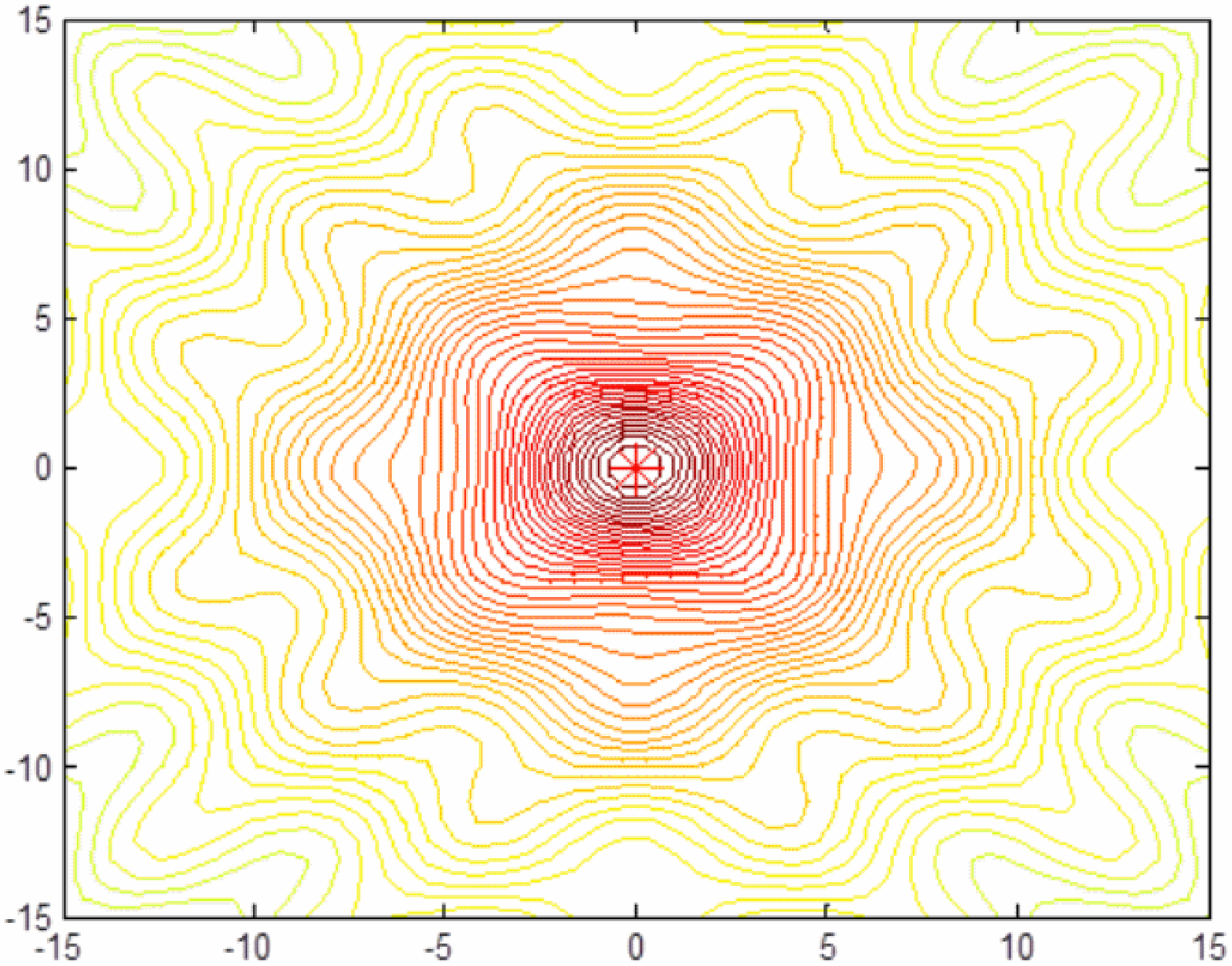}}
%\caption{(a)}
\end{minipage}
%\end{figure}%
%\begin{figure}[h]
\begin{minipage}{0.33\columnwidth}
\centerline{\includegraphics[width=1\columnwidth]{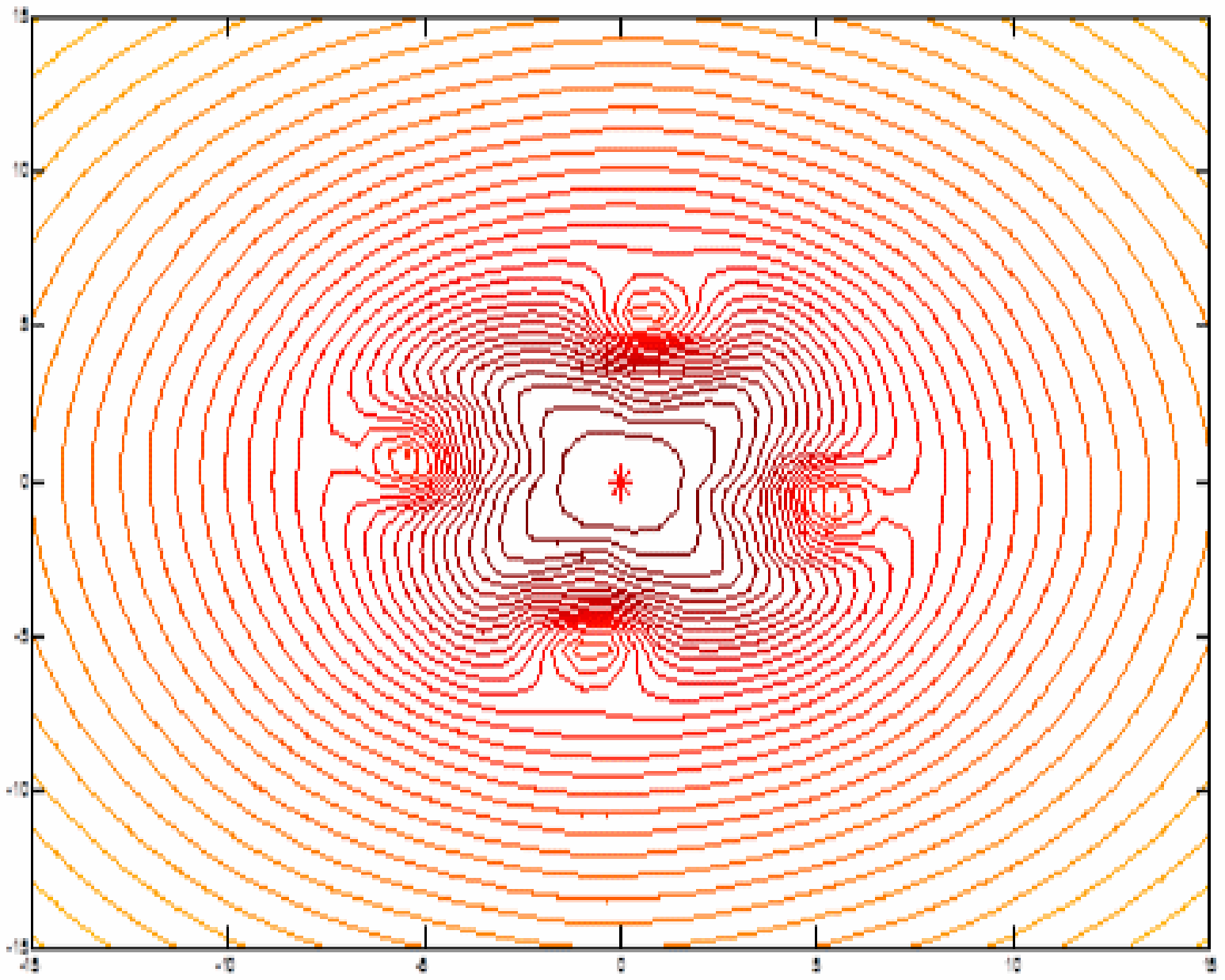}}
%\caption{(b)}
\end{minipage}%
\begin{minipage}{0.33\columnwidth}
\centerline{\includegraphics[width=1\columnwidth]{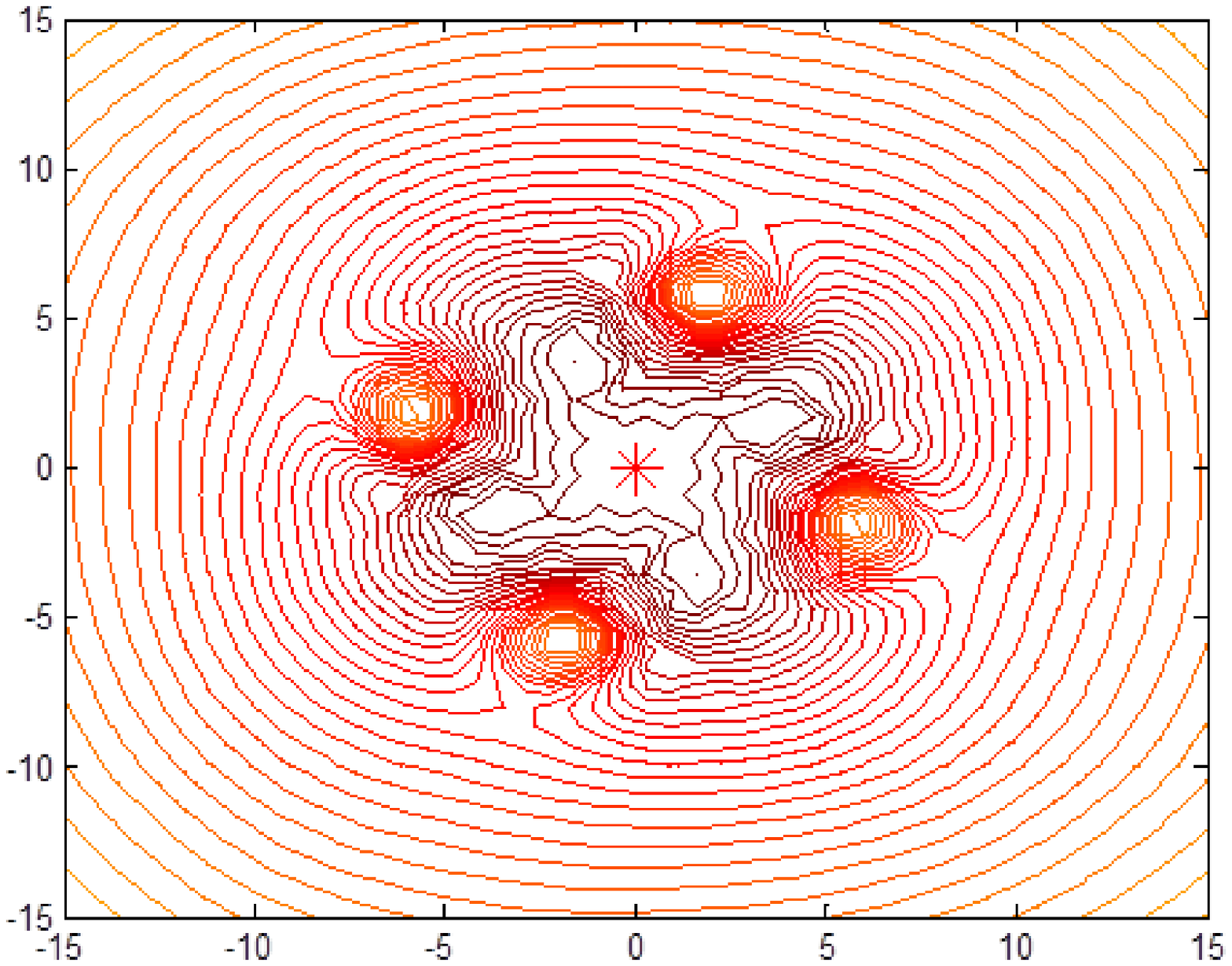}}
%\caption{(b)}
\end{minipage}%
\caption{Transformation of vortex for anticyclonic vorticity: $B_0=
10^{3}$, $B_0=7.5\cdot 10^{3}$ and $B_0=10^{4}$. $k=1.5$,
 $T\approx 24 \rm h$,  the horizontal scale is in kilometers}\label{instability2}
\end{figure}

 In the present paper, we studied numerically the deviation from the axial symmetry for
stationary solution having exponential decay rate. Namely, we use
initial data \eqref{ic_u}, \eqref{ic_pi}, with $k\ne 1$, where
parameter $k$ is a measure of deviation of velocity from the axial
symmetry. Here, we take $k=1.5$ and $b^*= B_0\sigma.$

We used the same numerical method as in \cite{RYH2012}. Namely, the
computations were made by a modified Lax-Wendroff scheme, the method
is second order accurate in both space and time  variables
\cite{Roache}, \cite{Lin}, \cite{Zhang}. The computations  were
performed on a $(400\times 400)$ uniform grid with the space step
$\Delta x = \Delta y = 0.64$ which corresponds to 12.8 km and the
time step $\Delta t = 0.0005$ which corresponds to 10 sec of the
real time.  We use the Neumann boundary condition set sufficiently
far from the vortex domain. Nevertheless, it is possible to use more
sophisticated non-reflecting boundary conditions (see \cite{Givoli})
and introduce an artificial viscosity to damp the oscillations
\cite{Reutskiy}.

The constants were chosen from the geophysical reasons, see
\cite{RYH2012}. Namely, the Coriolis parameter $l=7.3\times
10^{-5}\,{\rm s}^{-1}, $ that corresponds to the latitude
$30^\circ\,$ approximately, $c_0=0.1,\, R_0=10\,$(appropriate
dimension),
 $\gamma=\frac{9}{7}$ (in the
procedure of averaging over the height, the value of heat ratio for
air changes), $\sigma=10^{-9}\rm m^{-2}$.

The results of computations show that the solution blows up if the
value of $B_0 \sigma$ does not belong to the segment $(b_-, b_+)$,
$b_-\approx -1.31 \cdot 10^{-5}$, $b_+\approx 1.85 \cdot 10^{-5}$.
We note that the left boundary $b_-$ is sufficiently close to the
left boundary of stability domain for
$b^*=l\frac{1-\sqrt{2}}{2}\approx 1.51\cdot 10^{-5}$. The right
boundaries $b_+$ and $l\frac{1+\sqrt{2}}{2}$ are far from each
other. Let us note that real values of vorticity in the atmosphere
are sufficiently small. For example, in \cite{RYH2012}, we used
$B_0\sigma= - 10^{-6} \rm m^2/s. $

 Figs.\ref{instability1} and \ref{instability2} presents the level lines of pressure
after one day for the initial data with exponential decay
\eqref{ic_u}, \eqref{ic_pi}.

%for the initial data with power decay \eqref{ic_uq},\eqref{ic_piq}
%with $q=\frac12$.

\begin{remark}
We can see that the process of transition to breakdown is very
similar to the case of formation of four vortices of a smaller size.
Our graphs are similar to azimuthal Kelvin-Helmholtz instability,
known from experimental works, e.g.\cite{Flor2}.
\end{remark}

\begin{remark}
 The solution to the ODE system \eqref{m1}, \eqref{m2} can blow up
for certain initial data and the numerics confirms that this takes
place in the instability domains. For the system \eqref{main},
\eqref{pi} the blow up means  formation of infinite gradients of
solution, e.g. the shock waves.
\end{remark}

\begin{remark} If we choose $\Phi$ as a smooth function with compact
support, the initial data \eqref{U_formula}, \eqref{pi_formula} also
have compact support. So, we can apply the stability theorem from
\cite{Padula}, Sec.2.3.3, which says that the stationary solution of
the initial boundary-value problem with the Dirichlet boundary
condition is stable with respect to the class of smooth
perturbations.  Therefore, if for some range of parameters the
perturbed solution is unstable, then it does not keep smoothness for
all $t>0$.
\end{remark}

\section{Conclusion}\label{Sec4}

We examine both theoretically and numerically the nonlinear
stability of axisymmetric vortex in a compressible media.  As in our
previous works \cite{RYH2010}, \cite{RYH2012}, we consider two
dimensional barotropic model obtained by averaging over the height
of the primitive system of equations of the atmosphere dynamics. The
resulting model is the 2D model of compressible barotropic rotating
non-viscous medium. It can be used not only for studying the vortex
process in atmosphere, but also in plasma, etc. In contrast to other
models, where first the additional physically reasonable
simplifications are made, we deal with special classes of solutions
of the full system. This allows us to catch the complicated features
of the full model.

  We study how   rotation of the coordinate frame and
compressibility  affect the stability of vortex. It is known that
 the axisymmetric form of vortex is stable
with respect to asymmetric elliptical perturbations for the
solutions of the incompressible Euler equations in a fixed
coordinate frame (see \cite{Inviscid_damping} and references
therein). Both linear analysis and computations show that a
compressible vortex in a similar situation is basically instable
\cite{Chan}. We show that the situation is quite different for the
compressible vortex in a rotating coordinate frame. Namely, the
stability can take place only for a narrow range of parameters
characterized by the rotation of vortex with respect to the Coriolis
parameter. We prove that the motion is unstable if the parameter
$b_*/l$, characterizing the relations between the vorticity and the
Coriolis parameter, lies outside the segment $[(1-\sqrt{2})/2,
(1+\sqrt{2})/2]$. Using the analytical-numerical method, we checked
that the motion is unstable in a small neighborhood of boundary
points $(1-\sqrt{2})/2$, $(1-\sqrt{2})/2$, as well as in a small
neighborhood of the points
 $0$ and $1$, where the cyclonic motion changes to the  anticyclonic one.
 For points of the segment $[(1-\sqrt{2})/2, (1+\sqrt{2})/2]$,  up
to the third order terms in the expansion of the normal form at the
point of equilibrium, we checked numerically the sufficient
condition of the stability.

In particular, the results imply that a rotation can stabilize the
vortex. Similar results were obtained in \cite{Liu_Tadmor}, where
the authors conclude that the rotation prevents blow up in a
hydrodynamical model.

Furthermore, we study that the localized vortex with a linear
profile of velocity near its center numerically. We conclude that
for sufficiently small values of vorticity the range of parameters
guaranteing stability approximately corresponds  to the range of
parameters guaranteing stability of the vortex with a linear
profile.  The reason for the phenomenon is explained analytically.

As it was mentioned, in \cite{RYH2010}, \cite{RYH2012} we used the
class of solutions with the properties of linear profile velocity to
predict the trajectories of large geophysical vortices such as
tropical cyclones. As it follows from \cite{RYH2012}, \cite{R15},
the trajectory of localized vortex is not that far from the
trajectory of vortex with a linear profile of velocity, if the
divergence of velocity in this localized vortex is small in some
sense. Thus, in the stable cases, we can compute the trajectory of
vortices without much distortion even when the axial symmetry is
breaking.

%\begin{center}{ACKNOWLEDGMENTS}\end{center}

\ack The authors thank Professor Hung-Chi Kuo for a valuable
discussion. JLY was supported by the Academia Sinica in Taiwan for
Short-term Domestic Visiting Scholar. OSR was supported by
Mathematics Research Promotion Center, MOST of Taiwan. CKH was
supported by Grant MOST 104-2112-M-001-002.

\bigskip

%%%%%%%%%% Insert bibliography here %%%%%%%%%%%%%%

%Gundlach C 1999 {\it Liv. Rev. Rel.} 1994-4

%Alikakos N 1979 Lp bounds of solutions of reaction-diffusion
%equations Commun. Part. Differ. Equ. 4 827–68

%Sundu H, Azizi K, S\"ung\"u J Y and Yinelek N 2013 Properties of
%$D_{s2}^*(2573)$ charmed-strange tensor meson arXiv:1307.6058

%Dorman L I 1975 {\it Variations of Galactic Cosmic Rays} (Moscow:
%Moscow State University Press) p~103

\end{document}